**Lineage specific reductions in genome size in salamanders are associated with increased rates of mutation.**

John Herrick[1]* and Bianca Sclavi[2]*

*corresponding authors

1.  jhenryherrick@yahoo.fr

2.  LBPA, UMR 8113 du CNRS, ENS Cachan, Cachan, France 94235
sclavi@lbpa.ens-cachan.fr

*Abstract*
Very low levels of genetic diversity have been reported in vertebrates with large genomes, notably salamanders and lungfish [1-3]. Interpreting differences in heterozygosity, which reflects genetic diversity in a population, is complicated because levels of heterozygosity vary widely between conspecific populations, and correlate with many different physiological and demographic variables such as body size and effective population size. Here we return to the question of genetic variability in salamanders and report on the relationship between evolutionary rates and genome sizes in five different salamander families. We found that rates of evolution are exceptionally low in salamanders as a group. Evolutionary rates are as low as those reported for cartilaginous fish, which have the slowest rates recorded so far in vertebrates [4]. We also found that, independent of life history, salamanders with the smallest genomes (14 pg) are evolving at rates two to three times faster than salamanders with the largest genomes (>50 pg). After accounting for evolutionary duration, we conclude that more recently evolved species have correspondingly smaller genomes compared to older taxa and concomitantly higher rates of mutation and evolution.

## 1. Introduction

Low mutation rates are generally acknowledged to be required for the evolution of large genomes [5]. Hinegardner and Rosen first suggested in 1972 that fish with large genomes are evolving more slowly than fish with smaller genomes [6]. An investigation of evolutionary rates in lungfish (C-value 70 pg) likewise suggested that lungfish are evolving up to two fold more slowly than either frogs or mammals (C-value 3 pg) [7].

At the same time, larger genomes are more prone to mutation, and thus any increase in genome size is expected to impose a mutational hazard on the organism [8]. Studies in plants, fish and animals, for example, revealed a genome size correlation between extinction rates and species richness [9-12]. Taxa with large genomes experience higher rates of extinction, and tend to be less speciose. Together, these observations suggest that variations in mutation/substitution rates influence the mode and tempo of genome size evolution and rates of diversification in different plant and animal lineages.

In vertebrates, lineage specific mutation rate variation has been inconsistently associated with several different but interacting life history traits including body size and metabolic rate [13-15]. Generation time (GT), however, has been more consistently associated with variations in mutation rates, suggesting that errors in DNA replication during germ cell division are a primary source of mutation. A generation time effect, for example, has been proposed to account for the decrease in mutation rate resulting from DNA replication errors as the primate lineage evolved [16]. Low rates of molecular evolution in some acipenseriforme lineages have similarly been attributed to a generation time effect on mutation and substitution rates [17].

Generation times and developmental times are also strongly influenced by genome size [18]. In plethodontid salamanders, for example, embryonic developmental time is correlated with genome size [19]. The relationship between genome size and generation time is in part explained by the corresponding increase in nuclear and cell size, which frequently alters metabolic rate (nucleotypic effect) [20-22]. These observations suggest a possible relationship between genome size and mutation rate [23], and that GT influences mutation rates through secondary effects associated with genome size. An earlier proposal that genome size influences genetic variability generated some controversy [24, 25], yet how changes in genome size might interact with and affect mutation rates and rates of evolution remain unclear and relatively unexplored.

Amphibia, and urodela in particular, have been the subject of intense studies because of their evolutionary interest and their sensitivity to climate change [26, 27]. We have based our analysis on the wealth of sequence data, divergence time and genome size data from these studies to investigate the association between diversification rate and genome size in salamanders. Referring to published phylogenies [28-31], we compared the number of neutral substitutions that have occurred since the specific pairs within each lineage diverged. To do so, we measured substitutions at synonymous (dS) coding sites in the nuclear gene *rag1* in five different salamander families: Plethodontidae, Ambystomidae, Sirenidae, Proteidae and Cryptobranchidae. The latter four families are composed of entirely paedomorphic species.

## 2. Material and Methods

The nucleotide and amino acid sequences of orthologous *rag1* genes were obtained from GENBANK. Lineages were then selected from the obtained phylogenetic tree according to the availability of their C-values in the Animal Genome Size Database (www.genomesize.com) [32]. A MatLab script was implemented in order to search the data from the Animal Genome Size Database and then to search for the mean divergence times of species pairs with known C-values from the TimeTree website (www.timetree.org) [33].

Nucleotide sequences were codon aligned with the amino acid sequence using Pal2Nal [34] (www.bork.embl.de/pal2nal/) and CLUSTAL W in Mega 5 and finally refined by hand. Phylogenetic analyses were performed as follows: the best substitution model was chosen using MEGA 5 [35] and was used to generate a maximum likelihood tree. The resulting tree (Supplementary Figure 1 and legend for details) is consistent with previously published analyses.

Sequence divergences, the number of synonymous and non-synonymous substitutions per site, *pS* and *pN* were estimated in MEGA 5 [35] using the Nei-Gojobori and Kimura 2-parameter model with a gamma distribution value of 2. The *rag1* gene has the advantage that synonymous sequences are not saturated over the evolutionary distances considered here (not shown). Saturation was tested by plotting the proportion of synonymous nucleotide differences per synonymous site, pS, against sequence divergence for pairs of taxa [36].

Ten families of salamander were identified, with C-values ranging from 14 pg (*Desmognathus wrighti*) to 119 pg (*Necturus punctatus*). We measured dS between species pairs only within the family in order to control for clade specific effects. To correlate dS/Mya with a given genome size, we then selected species pairs within each lineage that had similar genome sizes (± 10% of genome size). Each point in the plots thus represents the dS/Mya versus the average value of the two respective genome sizes (error bars of 10% in genome size).

## 3. Results

*3.1 Genome size variation in fish, frogs and Salamanders*

Earlier studies in plants, fish and animals revealed an association between genome size, extinction rates and species richness [9-12]. The association between genome size and species richness becomes especially apparent in groups with genome sizes larger than 5 pg in amniotes and 14 pg in plants [10, 11]. Here we examined the distribution of species as a function of genome size in three distantly related groups: fish, frogs and salamanders. The genome size of each species was obtained from the Animal Genome Size Database [32].

Figure 1 shows that the distribution of genome sizes is significantly different for each group. Fish typically have genome sizes of one to two picograms (pg), mammals (not shown) have genome sizes typically between two to four picograms, frogs have a genome size of four to six picograms while salamanders range from 14 to 50 pg and larger. Genome size distributions in salamanders, however, appear to be more bimodal with a peak at 20 pg and another at 40 pg.

The 20 pg peak corresponds mainly to Salamandridae, Ambystomidae and Hemidactyinii, the 40 pg peak to the genera *Bolitoglossa* and *Plethodon* (the latter spanning a wide range including 20 pg). Salamanders, as previously reported, therefore have a genome size that is on average ten times larger than other vertebrates, with the exception of lungfish [37]. We propose that non-random evolutionary constraints restrict the range of genome size in different vertebrate groups. Accordingly, genome size reflects the balance between the forces of genetic drift and selection acting at the cellular, developmental and environmental levels.

*3.2 Low but heterogeneous substitution rates in the rag1 gene in Plethodontidae*

The results presented in Figure 2A show the distribution of evolutionary rates among the Desmognathinae, Plethodontini and *Hydromantes*. The Desmognathinae, which have the smallest genomes, are evolving faster as a group than either the Plethodontini or *Hydromantes*. Figure 2A also reveals that within the *Plethodon* the lower limit for mutation rates for species pairs with smaller genomes is higher than for species pairs with larger genomes. A similar but less steep trend can be seen at the higher limit of the distribution reflecting an increase in the variance of evolutionary rates in larger genomes.

*3.3 Evolutionary rates within the genus Bolitoglossa decrease with increasing genome size.*

The genus *Bolitoglossa* is one of the most diverse genera of salamander, and comprises about 120 species. This genus radiated during the early Miocene in Central and South America [38]. Genome size ranges from 42 pg to 68 pg (average: 58.7 pg). In contrast, the genus *Plethodon*, which originated about 35 MYA [31], has a genome size that ranges from 18 pg to 49 pg (average: 20 pg). Although average genome sizes are significantly different between the two genera, the overall range of genome size (30 pg) is similar in both, as is the range of evolutionary rates (0.001 to 0.007 dS/MYA). *Bolitoglossa*, however, exhibits a much clearer and more pronounced negative correlation between genome size and evolutionary rates than is found in *Plethodon*. In both cases however a mutation rate of 0.001 to 0.002 dS/MYA appears to represent a lower limit in larger genomes. We conclude that within *Bolitoglossa* higher rates of mutation and evolution are associated with smaller genomes (Figure 2C).

*3.4 Evolutionary rates in obligate paedomorphs are negatively correlated with genome size*

Figure 3 shows the distribution of evolutionary rates versus genome size for aquatic paedomorphs. These species consistently have the lowest levels of genetic diversity and the largest genomes among salamanders. Life history and ecological factors, such as neoteny and a narrow niche, have been invoked to explain the low levels of diversity and large genomes in paedomorphs. *Eurycea* (Plethodontidae), however, appears to be an exception. *Eurycea* are acquatic paedomorphs, but have exceptionally high rates of evolution among all salamanders and significantly smaller genome sizes than other paedomorphs. We conclude that in paedomorphs higher evolutionary rates are associated with smaller genomes independent of life history or niche width.

The correlation between genome size and evolutionary rates is even more apparent in Figure 3: in paedomorphs, evolutionary rates accelerate as genome size decreases in a manner similar to that found in the *Bolitoglossa* genus (regression slope = -2.6). Evolutionary rates in paedomorphs decline to a minimum of about 0.001 dS/Mya, which are among the slowest evolving rates reported for vertebrates. The Proteidae (*N. maculoso* and *N. punctatus*), however, appear to be an exception to the trend. Rates of evolution are higher despite these species having the largest genomes, an observation that is consistent with the finding that the Proteidae have higher levels of heterozygosity compared to the other paedomorphs [39].

Allowing for this exception, a clear negative trend nevertheless emerges between genome size and evolutionary rates: *Eurycea* (20 pg) is evolving faster as a group than Ambystomidae (25 pg), which are evolving faster than the Chryptobranchidae (50 pg), the Sirenidae (55 pg) and Amphiumidae (75 pg). The latter three families appear to be at the evolutionary lower limit of permissible mutation rates.

*3.5 Smaller genome sizes in more recently evolved urodels*

Figure 4 shows the average genome size for the species of a given genus and the time of origin (evolutionary duration). The time of origin was obtained from Marjanovic and Laurin [29]. As previously reported for neotenes [40], the trend in decreasing genome size in clades of more recent origin is thus extended to and confirmed for terrestrial and direct developing salamanders. *Pleurodeles* and Desmognathinae, for example, are the most recently evolved salamanders examined here, and they exhibit the smallest genome sizes.

Interestingly, *Eurycea*, with the smallest genome size among paedomorphs, is also more recently evolved. Earlier, it was suggested that neotene salamanders have large genomes due to their fluctuating aquatic environments, which are expected to result in small effective population sizes [41]. The smaller, rapidly evolving genome of *Eurycea*, however, suggests that life history, although important, is not predominantly responsible for the evolution of large genomes in paedomorphic salamanders.

These observations suggest that either genome sizes have expanded at a constant rate since the origin of a family, thus resulting in larger genomes in older families; or, conversely, speciation events have resulted in increasingly smaller genome sizes; or both processes might be impacting the mode and tempo of genome size evolution. We conclude that smaller genome sizes are associated with more recently evolved species across salamander taxa, and are concomitantly associated with higher mutation rates and rates of evolution.

4. Discussion

Little evidence currently exists that genome size influences either mutation rates or rates of evolution. The results presented here suggest that rates of evolution tend to increase as genome size decreases. Among the Plethodontidae, the Desmognathinae are evolving faster than species with larger genomes, such as the *Hydromantes*. The predominantly terrestrial

Plethodontidae, however, exhibit the largest variance in the correlation between genome size and evolutionary rates, indicating that a larger niche breadth has significantly influenced evolutionary rates in this particularly speciose family of salamander. This observation is consistent with the niche-width variation hypothesis [39], which proposes that greater habitat heterogeneity is associated with greater genetic diversity.

The negative correlation found here is most strikingly apparent within the single genus of *Bolitoglossa* (Figure 2C): genome size and evolutionary rate are inversely correlated even over relatively small differences in genome size (<10 pg). The correlation is much stronger than among the *Plethodon* (20 pg) and Desmognathinae (16 pg) of the same family (Figure 2B). Although the larger variance in evolutionary rates among the Plethodontidae weakens the correlation significantly (Figure 2B), there appears to be a lower limit in the rate of mutation that is dependent on genome size: minimum observed mutation rates clearly increase as genome size decreases.

Similarly, among paedomorphs, a clear negative correlation exists between genome size and rate of evolution (Figure 3): Hemidactylini (20 pg), from the Plethoditidae, are evolving faster than Ambistomidae (35 pg), which in turn are evolving faster than the Sirenidae and Amphiumidae (>50 pg). The Proteidae, which have the largest genomes examined here, appear to be an exception, although they too are more slowly evolving than Hemidactylini. Hence, the observations on the aquatic paedomorphs are consistent with the terrestrial families: species with the smallest genomes tend to evolve faster than species with the largest genomes.

Earlier studies reported an apparent correlation between nuclear DNA content and evolutionary duration in salamanders (Figure 4) [40]. The authors interpreted the trend as evidence that genome size increases depending on how long a species has been an obligate neotene, and proposed that the rate of junk DNA accumulation could be used as a possible second molecular clock. Alternatively, the trend reproduced here might suggest that reductions in genome size accompany evolution and speciation in salamanders consistent with observations made in other organisms [12]. We have shown additionally that the apparent trend of smaller genomes in more recently evolved taxa is associated with increased rates of mutation/diversification, as observed in other vertebrates and plants [42, 43]. These changes are suggestive of a re-patterning of non-coding DNA and its reorganization during the process of speciation.

Why might mutation rates as assessed here tend to increase as genome size decreases? We propose that larger amounts of nuclear DNA could act during the cell cycle and development to facilitate adaptive improvements in the efficiencies of DNA replication fidelity and DNA damage repair in coding DNA (for a review see [44]), while at the same time the forces of genetic drift might impose a lower limit on mutation rates (0.001 to 0.002 dS/MYA for vertebrates) [45, 46]. Consequently, larger genomes are expected to have lower mutation rates in coding DNA and correspondingly higher mutation rates in non-coding DNA, a prediction that we are currently investigating. This proposal is consistent with a nucleotypic/generation time effect on evolutionary rates [47], according to which smaller genomes are often associated with more rounds of germline genome duplications, which results in an accumulation of errors, and have correspondingly less time in the cell cycle to repair errors and substitutions.

Recently, Sun *et al*. revealed that low rates of DNA loss contribute to genomic gigantism in salamanders [48]. In contrast, we suggest that although salamander genomes tend to be large as a result of bias against DNA loss, genome size reductions nevertheless might accompany salamander evolution (Figure 4 and Supplementary Figure 1), a hypothesis that remains, however, to be verified. Together with other studies on genome evolution and speciation [49, 50], the findings presented here suggest that genome size influences, directly or indirectly, rates of evolution and speciation, and that the inverse relationship between genome size and evolutionary rates might represent a more general evolutionary principle.

## 5. Acknowledgements

The authors would like to thank Michel Laurin for his helpful comments on the manuscript. BS is supported by a grant from HFSP (RGY0079). JH benefitted from technical support from John Bechhoefer's lab, Physics Department, Simon Fraser University.

**Figure legends**

**Figure 1. Distribution of C-values in fish, frogs and salamanders.** A) Frogs (anura) exhibit a 14 fold range in genome size between one and fourteen picograms (pg) that approximates a Gaussian distribution. B) Ray-finned fish exhibit a significantly narrower range of genome size between 0.4 to 5 pg, with a mode at 1 pg. C) Salamanders (urodela) exhibit a more complex distribution and display two clear peaks. The peaks are centered at 27 and 38 pg, and correspond to distinct salamander families. The C-values in each peak fit a Gaussian distribution. D) Cartilaginous fish display a broader range of genome size (1 to 18 pg) with a mode at 4 pg. The mode value is 4X that of ray finned fish. Both distributions, however, approximate a log normal distribution of C-values, suggesting a different mode of genome evolution compared to tetrapods [51].

**Figure 2: Evolutionary rates in metamorphosizing and direct developing Plethodontidae.** A) Combined datasets of Plethodontidae reveals a negative correlation between evolutionary rates (dS/MYA) and genome size (picograms) across the different genera examined here (slope = -0.268; SE = 0.19). B) *Desmognathus* and *Plethodon* rates of evolution. *Desmognathus* species are both metamorphic and direct developing and represent some of the smallest genomes among salamanders. *Plethodon* (direct devolpment) display a large variance in dS/MYA, suggesting adaptations to more heterogenous, terrestrial niches. Note that in *Plethodon*, the slowest observed rate of evolution for a given C-value increases with genome size between 20 and 32 pg. Similarly, the fastest observed rates for a given C-value decreases with increasing genome size (slope = -0.84; SE = 0.45). C) *Bolitioglossa* evolutionary rates. The negative correlation between evolutionary rates and genome size is most apparent in this genus, although the variance appears to increase with genome size: the smaller the genome the lower the variance in rates. (slope = -2.64; SE = 1.47)

**Figure 3: Evolutionary rates in aquatic paedomorphs.** The *Eurycea* species examined here are evolving up to 10X faster than either the *Andrias* species or species belonging to the Sirenidae family. Both sets of species have genome sizes that are twice as large as the *Eurycea*. *Andrias* and *Amphiuma* species are the slowest evolving species observed here. Species belonging to the Proteidae (*Necturus*), in contrast, are evolving faster than the latter two despite having the largest genomes in this dataset, an observation that is consistent with a higher level of heterozygosity compared to Chryptobranchidae, Sirenidae and the Abystomidae (slope = -2.26; SE = 0.32).

**Figure 4: Negative correlation between average C-value and evolutionary duration (time since divergence).** The Animal Genome Size Database [32] was searched for those families with published time of origin (data from Marjanovic and Laurin [29]). The regression line reveals a clear relationship between decreasing genome size and more recently evolved species: genome size is progressively and consistently smaller in more recently evolved species. The slope is approximately 1 pg/MYA, which differs from the previously reported slope of 0.629 pg/MYA found for paedomorphic families of salamander [40].

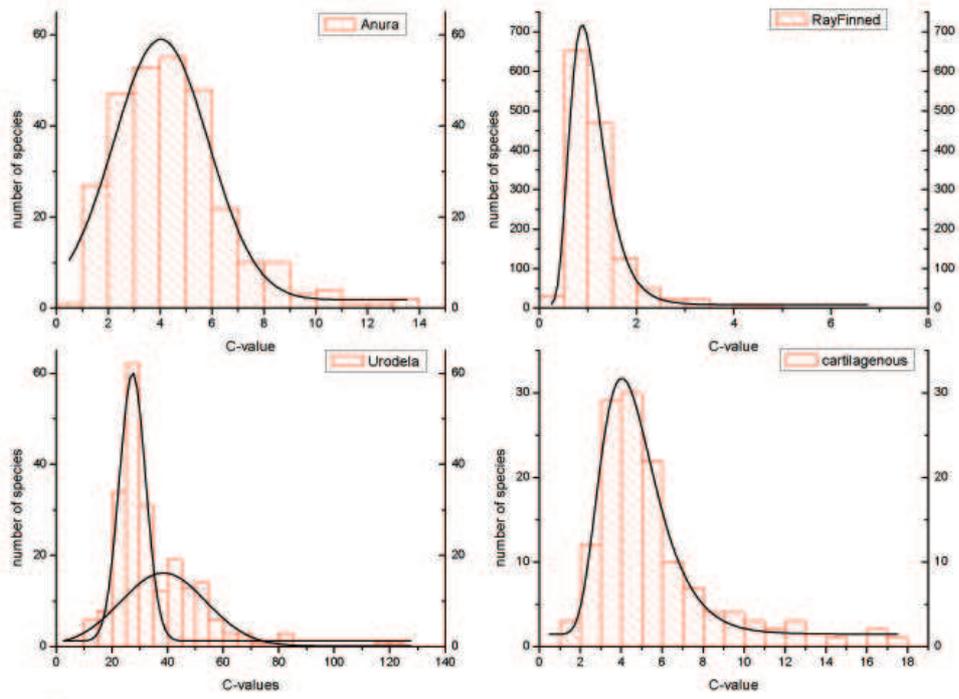

Figure 1

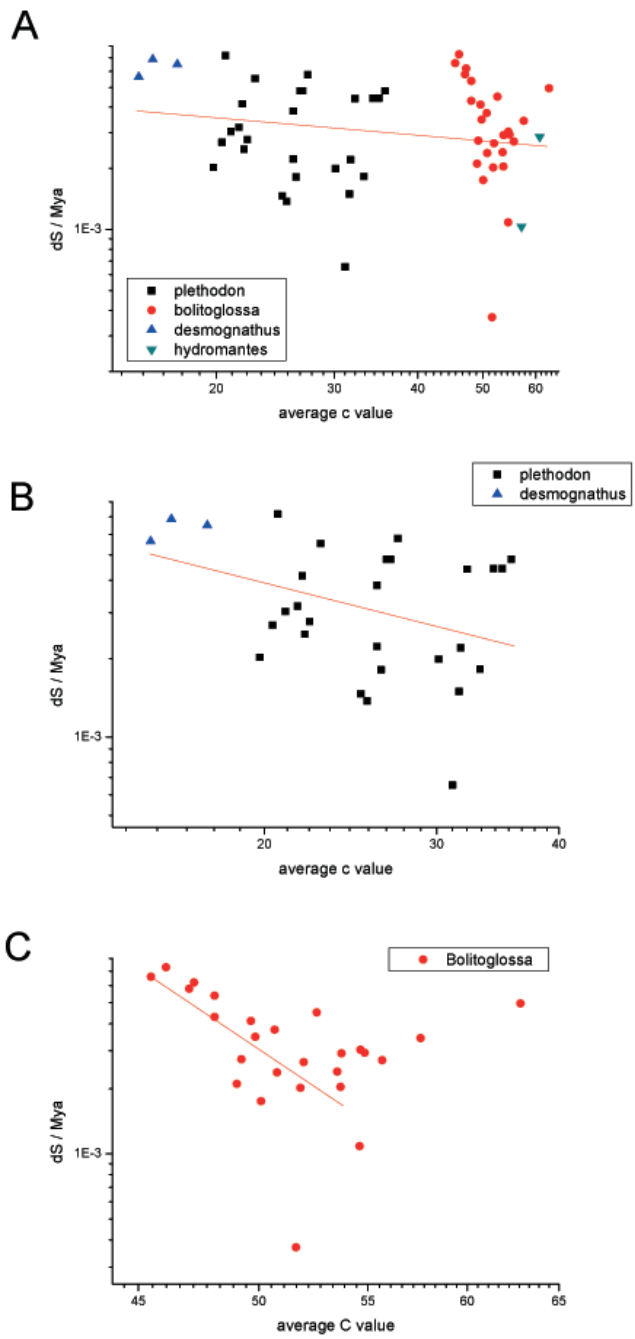

Figure 2

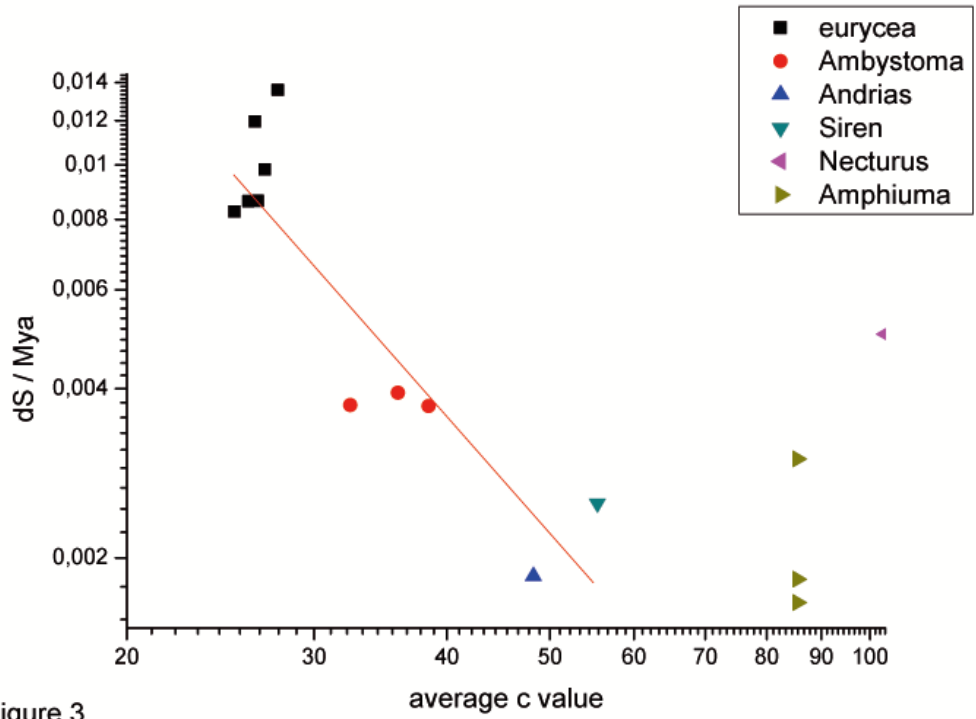

Figure 3

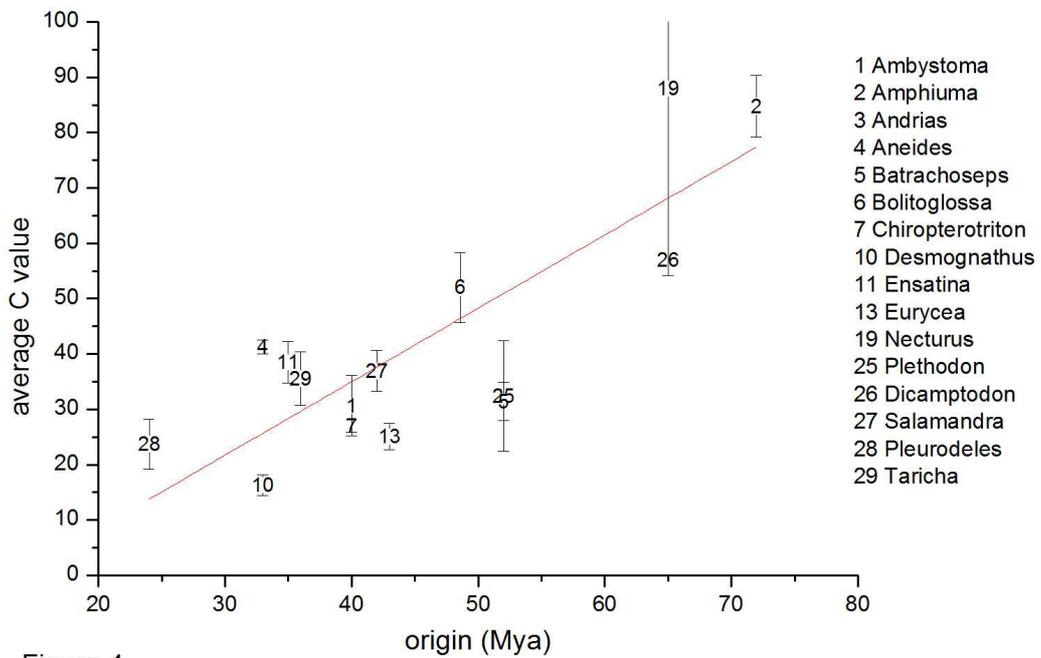

Figure 4

1 Ambystoma
2 Amphiuma
3 Andrias
4 Aneides
5 Batrachoseps
6 Bolitoglossa
7 Chiropterotriton
10 Desmognathus
11 Ensatina
13 Eurycea
19 Necturus
25 Plethodon
26 Dicamptodon
27 Salamandra
28 Pleurodeles
29 Taricha

**Supplementary materials**

**Figure 1S: Molecular Phylogenetic anaylsis by Maximum Likelihood method**
The evolutionary history was inferred by using the Maximum Likelihood method based on the Kimura 2-parameter model [1]. The tree with the highest log likelihood (-2708.4283) is shown. The percentage of trees in which the associated taxa clustered together is shown next to the branches. Initial tree(s) for the heuristic search were obtained by applying the Neighbor-Joining method to a matrix of pairwise distances estimated using the Maximum Composite Likelihood (MCL) approach. A discrete Gamma distribution was used to model evolutionary rate differences among sites (5 categories (+*G*, parameter = 0.5169)). The rate variation model allowed for some sites to be evolutionarily invariable ([+*I*], 50.9028% sites). The tree is drawn to scale, with branch lengths measured in the number of substitutions per site. The analysis involved 69 nucleotide sequences. All positions with less than 0% site coverage were eliminated. That is, fewer than 100% alignment gaps, missing data, and ambiguous bases were allowed at any position. There were a total of 271 positions in the final dataset. Evolutionary analyses were conducted in MEGA5 [2]. The C value from the Animal Genome Size Database [3] is indicated to the right of each species name, the minimum and maximum is shown when relevant.

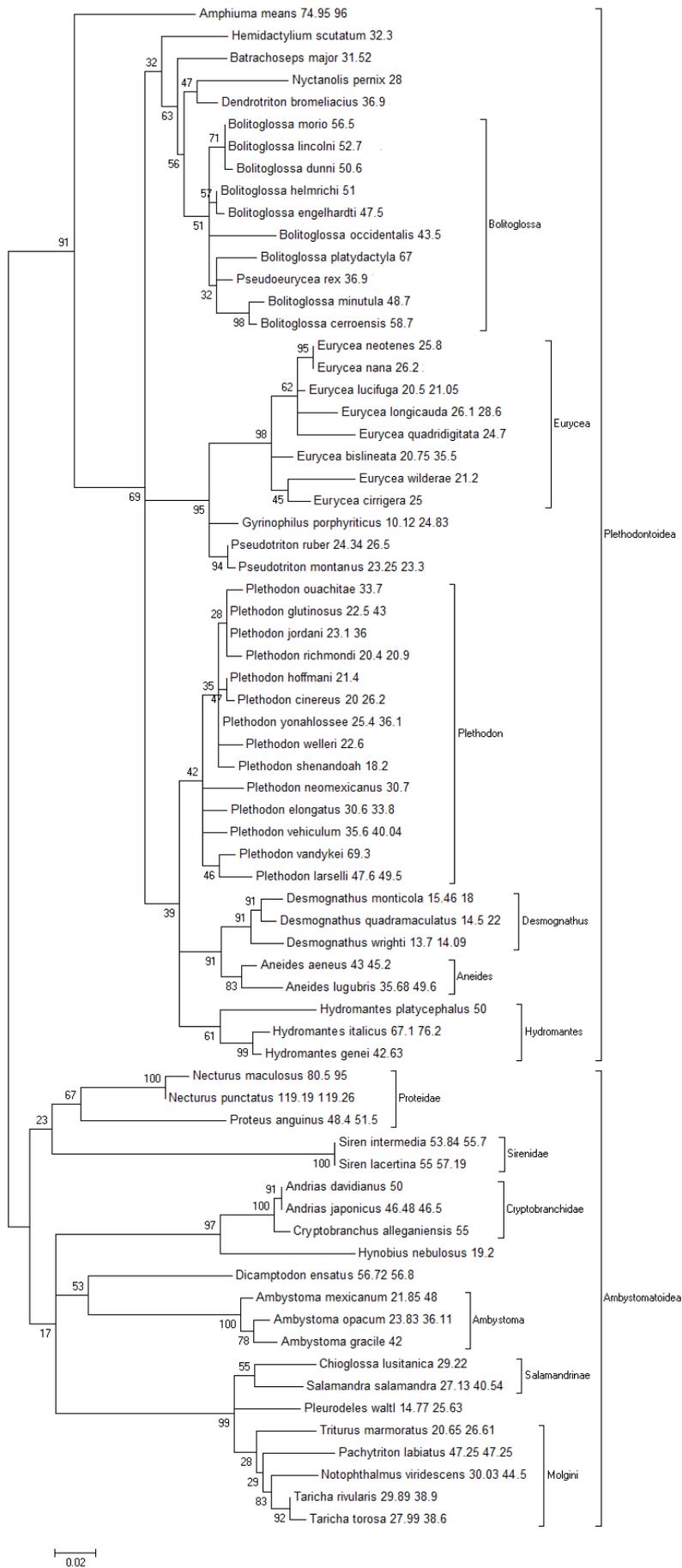